\documentclass[conference]{IEEEtran}
\IEEEoverridecommandlockouts
\usepackage{cite}
\usepackage{graphicx,graphics,epsfig,subfigure,times,bm,bbm,amssymb,amsmath,amsfonts,amsthm,mathrsfs,MnSymbol,xcolor}
\usepackage{gensymb}
\usepackage[matrix,frame,arrow]{xypic}
\usepackage[pdfstartview=FitH]{hyperref}
\usepackage{subfigure}
\usepackage{braket}
\usepackage{enumerate}
\usepackage{tikz}
\usetikzlibrary{calc}
\usepackage{float}

\usepackage{amsmath,amssymb,amsfonts}
\usepackage{algorithmic}
\usepackage{graphicx}
\usepackage{textcomp}
\usepackage{xcolor}
\def\BibTeX{{\rm B\kern-.05em{\sc i\kern-.025em b}\kern-.08em
    T\kern-.1667em\lower.7ex\hbox{E}\kern-.125emX}}
\begin{document}

\title{Multipath entanglement purification strategies for quantum networks\\
\thanks{DST Govt. of India, SERB grant MTR/2022/000389; IITB TRUST Labs grant DO/2023-SBST002-007; IITB seed funding}
}

\author{\IEEEauthorblockN{Md Sohel Mondal}
\IEEEauthorblockA{\textit{Department of Physics and CoE-QuICST} \\
\textit{Indian Institute of Technology Bombay}\\
Mumbai 400076, India \\
mondalsohelmd@gmail.com}
\and
\IEEEauthorblockN{Siddhartha Santra}
\IEEEauthorblockA{\textit{Department of Physics and CoE-QuICST} \\
\textit{Indian Institute of Technology Bombay}\\
Mumbai 400076, India\\
santra@iitb.ac.in}}
\maketitle
\begin{abstract}
In quantum networks multipath entanglement purification (MEP) between a pair of source-destination nodes can substantially strengthen their entanglement connection. An efficient MEP strategy can therefore increase the size of the network region where bipartite entanglement based quantum information processing tasks can be implemented. Here, we analyse MEP in a general model of a quantum network and obtain design criteria for efficient MEP strategies. Further, by simulating two different MEP strategies, based on these criteria, on different underlying network topologies we explore how the topology determines the effectiveness of a fixed MEP strategy. Finally, we show that a careful choice of MEP strategy can make the entanglement connection strength between source-destination network nodes effectively independent of its topology. Our results can therefore provide a useful guide for the design of quantum networks and entanglement distribution protocols.
\end{abstract}
\begin{IEEEkeywords}
quantum networks, entanglement routing, entanglement purification
\end{IEEEkeywords}
\section{Introduction}
Quantum information processing (QIP) tasks in a quantum network \cite{qn_pirandola,qn_wehner,qn_simon} can be be performed between two network nodes provided the entanglement of the state shared by them is above the given task's entanglement thresholds. In general, however, the entanglement distributed between any two network nodes decreases with the distance between them. This limits the effective size of the network regions over which the task can be performed. Therefore, as quantum networks mature, it is important to leverage the ability to perform sophisticated entanglement manipulation operations such as entanglement purification \cite{purification_benett,purification_deutsch} of the states distributed along multiple network paths to increase the effective network size relevant for QIP applications.

 Multipath entanglement purification is a network protocol that exploits the complex network structure of a quantum network and entanglement purification to strengthen the entanglement connection between network nodes measured by the concurrence of the final shared entangled state. In large-scale quantum networks, the complex network structure provides multiple alternate paths between a given pair of source-destination ({\it S-D}) nodes \cite{complex_qn}, each of which can establish an entangled quantum state distributed between the said nodes. MEP utilises these entangled states in the standard entanglement purification process to probabilistically obtain a single output state with a higher value of its concurrence.
 
A specific MEP strategy corresponds to the particular order in which various available alternate paths between the same pair of network nodes are pairwise and sequentially purified resulting in an output state with entanglement that depends on the order in general. This is because the different paths have varying lengths, measured in terms of the number of edges or hops, and the concurrence of the state distributed along a network path depends on its length and the statistical distribution of edge-entanglement values. An efficient MEP strategy seeks to maximize the output concurrence by optimising the order in which the paths are purified. Such an optimisation relies on knowledge of the global network topology, that is the network graph, and the distribution of the edge-concurrence values. 

In this work, we obtain design criteria for efficient MEP by
\begin{enumerate}
 \item Showing analytically that it is useful only when the lengths of the network paths that are used for purification are comparable in an appropriate sense, discussed in Sec. (\ref{subsec:comparable_paths}).
\item A few network paths are sufficient for MEP to approach the asymptotic value of concurrence for networks with high mean edge-concurrence in Sec. (\ref{subsec:mep_strategy}).
\item Finally, we show that the order in which the states along different alternate paths are purified determines the final output concurrence and that the sequence of purification steps in which the state along the shortest-path is included in the end typically results in the highest final output concurrence, also in Sec. (\ref{subsec:mep_strategy}).
 \end{enumerate}

Since the topology of a quantum network determines the number, availability and lengths of distinct paths between arbitrary {\it S-D} node pairs, it is clear that the effectiveness of any MEP strategy depends on the network topology. Networks based on regular lattices such as the square, triangular, hexagonal etc. have a fixed degree for each node and the numbers and lengths of distinct available paths between any pair of nodes can be deterministically obtained. On the other hand for random and scale-free (Barabasi-Albert) networks \cite{steen_complex_networks} these quantities differ for each node pair and can only be statistically specified. Numerical simulation of MEP strategies on all these types of networks reveal the effect of increasing node degree, path lengths and network topologies on the effectiveness of MEP. 

We discuss these results in the following starting from a description of the quantumn network model, network operations, network topologies in Sec. (\ref{sec_two}), multipath purification and strategies in Sec. (\ref{sec_three}), numerical simulation results in Sec. (\ref{sec_four}) and concluding with a discussion in Sec. (\ref{sec_five}).

\section{Background: entanglement distribution and topology of quantum networks}
\label{sec_two}
In this section we describe a general mathematical model for quantum networks including the form of the network states considered. Then we describe the basic network operations of entanglement generation, swapping and purification that transforms short-range to long-range entanglement in the network. Third, we describe some possible network topologies based on regular lattices as well as random and scale-free graphs. Finally, we describe multipath entanglement purification on the network graph. 

\subsection{Network Model}
We consider a quantum network based on an underlying graph $\{G(V,E),~\text{Dist}(c_e)\}$, where $V$ and $E$ denotes the set of nodes and edges in the network respectively; with $0\leq c_e\leq 1$ a random variable representing the concurrence of the entangled state generated along the edges $e\in E$ of the network. In general, the random variable, $c_e$, has a statistical distribution, $\text{Dist}(c_e)$, over the set, $E$, of the network edges. In homogenous networks all edges have the same value of the concurrence, that is, $c_e=c~\forall e\in E$. Whereas, for inhomogenous networks, we take the distribution of $c_e$ in terms of three quantities, its maximum, mean and minimum as follows,
\begin{align}
    &\text{Max}(c_e)=1-a\delta,~ 0\leq a<1 \nonumber \\
    &\text{Mean}(c_e)=1-\delta,\nonumber \\
    &\text{Min}(c_e)=1-b\delta,~ 1\leq b\leq 1/\delta
    \label{conc_dist}
\end{align}
where, $0\leq\delta\leq 1$ is a measure of the average edge-concurrence. For, $\delta\to0$ one has a quantum network with the edge-concurrences that are close to the maximum value of 1. The quantity $(b-a)$ is a measure of the homogeneity of the network. It becomes zero for homogeneous networks with increasing values denoting a higher spread of edge-concurrence values. In our numerical simulations we consider inhomogenous networks of $|V|=10^4$-nodes and $5\times10^4$-edges with, $\text{Max}(c_e)=0.99, \text{Mean}(c_e)=0.98$ and $\text{Min}(c_e)=0.97$. That is, the considered networks have a high average edge-concurrence value with relatively small variance over the edges.

Physically, the nodes $v\in V$ represent spatially separated (potentially over geographical distances) quantum systems whereas the edges are the quantum channels between two network nodes. We assume the form of the entangled state over any edge to be described by an Isotropic state which has the form,
\begin{align}
    \rho(q_e)=(1-q_e)\ket{\phi^+}\bra{\phi^+}+q_e\frac{\mathbf{1_4}}{4}
    \label{edge_state}
\end{align}
Here, $\ket{\phi^+}=(\ket{00}+\ket{11})/\sqrt{2}$ and $0\leq q_e\leq 1$. Isotropic states are a one-parameter family of mixed states that are routinely considered in analyses of quantum networking due to the simplicity of their form and operational interpretation. The state in Eq. (\ref{edge_state}) can be understood as a convex mixture of the maximally entangled pure state, $\ket{\phi^+}$, produced by an entanglement source with white-noise $\mathbf{1}/4$ introduced by the quantum channel. 

The concurrence of the states in Eq.~\ref{edge_state} can be expressed in terms of the state parameter $q_e$ as $c_e=\text{Max}(0,1-\frac{3}{2}q_e)$ which implies that the state is entangled for $q_e<2/3$. This means that for sufficiently low values of the noise parameter, $q_e$, along a network edge the end nodes share an entangled quantum state. Typically, however, when entanglement is distributed over network paths comprising multiple edges by swapping the entanglement over adjacent edges the noise parameter excceds this value and the concurrence goes to zero. Our analysis therefore focuses on {\it S-D} node pairs within network regions where statistically non-zero values of the concurrence may be expected.

\subsection{Quantum network operations}
In a quantum network the distribution of entangled states between the network nodes involves three probabilistic entanglement manipulation operations: entanglement generation (EG), entanglement swapping (ES) and entanglement purification (EP). In the process of EG, entangled qubits are created and they are shared between the neighbouring nodes, hence establishing short-range entanglement in the form of states $\rho(q_e)$ between nodes connected by an edge as given by the network graph, $G(V,E)$. Then, entanglement swapping extends the short-range entanglement to long-range entanglement in the network by performing a measurement on half of each pair of the two entangled state that share a node. Since, the first two operations may degrade the quality of entanglement established between two end nodes, the third network operation addresses this issue and increases the quality of entanglement by purifying the long-range entangled states shared via different available paths between a particular pair of network nodes.

\subsection{Topologies of network graphs}
Since in subsequent sections we discuss MEP for various network graphs, here we describe different topologies for such quantum networks. We focus on five distinct network topologies : Random network (RN), Barabasi-Albert network (BAN), Triangular lattice network (TLN), Square lattice network (SLN) and Hexagonal lattice network (HLN).

A RN is a type of network  in which the edges connecting the nodes are created at random according to some probability. RNs are often used as a reference to compare and analyse more complex networks. RNs show the small-world property which implies that any two nodes of the network may be connected by a path of adjoining edges which is not too long ($\lesssim \log(|V|)$) as a function of the number of nodes of the graph, $|V|$. The degree distribution of RNs is Poissonian, however, most real world networks deviate from this. Real world networks often tend to have a scale-free network structure where a few nodes have a very high degree and most of the other nodes have relatively low degrees. Scale free networks, also known as Barabasi-Albert networks, have a power-law degree distribution. Such networks have the ultra small-world property where the distance between an arbitrary pair of nodes is statistically very small, $\lesssim\log\log(|V|)$, as a function of the network size. The scaling of these expected maximum path lengths between network nodes have important design implications for quantum networks in terms of the required number of entanglement swapping operations and fidelity of the states along the network edges \cite{qn_topography}.

We also consider other three networks topologies based on regular lattices. TLN is the network based on an underlying triangular lattice grid where maximum of the nodes, except the ones at the boundaries, have degree 6. SLN and HLN are the networks based on square and hexagonal grids respectively. The degree of most of the nodes, that are not at the boundaries, in SLN and HLN are 4 and 3 respectively.

\begin{figure}
    \centering
    \includegraphics[width=\linewidth]{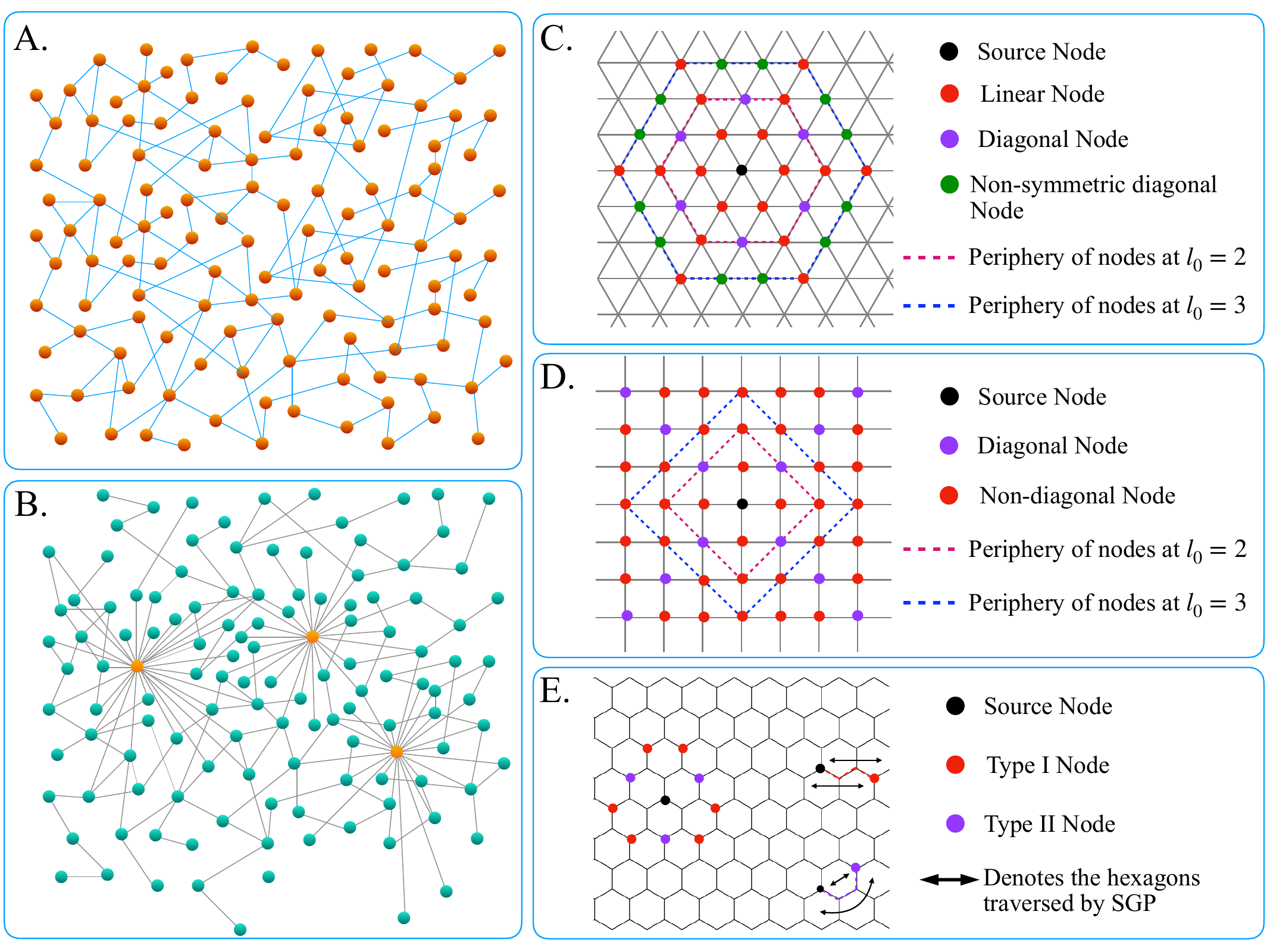}
    \caption{(A) Random network. (B) Scale-free network. (C) Triangular lattice network. Black node is the source node, $S$, and there are three different types of destination nodes, $D$, based on their path lengths from $S$. Red nodes are linear $D$ nodes, purple nodes are diagonal $D$ nodes and green nodes are non-symmetric diagonal nodes which can be considered as the diagonal point of $S$ on a parallelogram. (D) Square lattice network. In SLN we get diagonal and non-diagonal $D$ nodes, drawn in purple and  red color respectively. (E) Hexagonal lattice network. In HLN we have two types of $D$ nodes, type I and type II. This distinction is based on the equality of the number of hexagons traversed by the SGP on both the sides, for type I nodes the SGP traverses equal number of hexagons on both the sides of it but for type II nodes the number is different.}
    \label{fig:network_topologies}
\end{figure}

\begin{figure}[H]
    \centering
    \includegraphics[width=\linewidth]{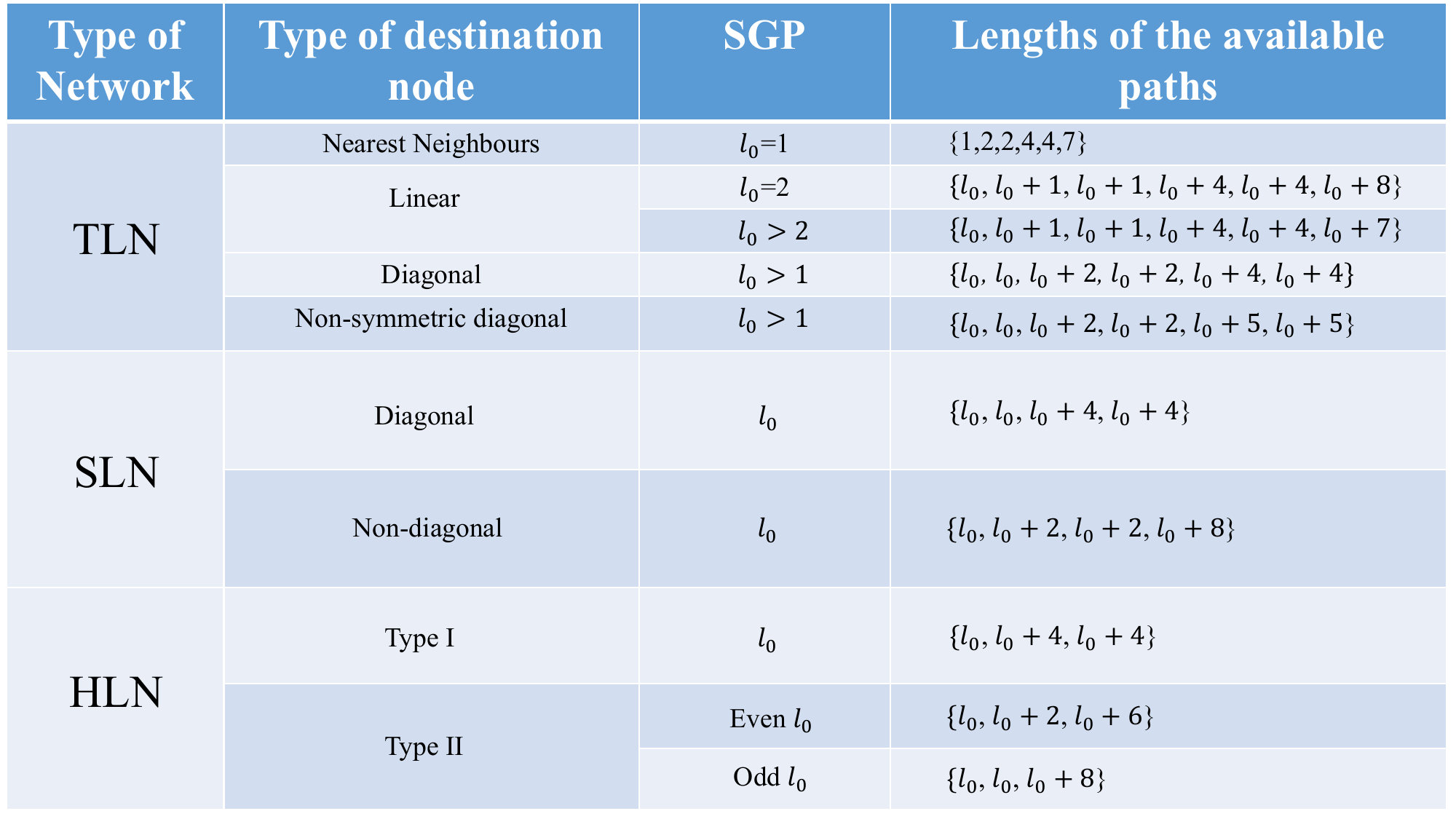}
    \caption{Lengths of the available paths between different types of {\it S-D} node pairs in the three regular network topologies, triangular (TLN), square (SLN) and hexagonal (HLN). Depending on the relative positions of the {\it S-D} pair shown in the second column and the characteristics of the shortest graph path (SGP) in the third column, the paths lengths of the distinct alternate paths available for the three topologies are listed in the last column.}
    \label{fig:lengths_table}
\end{figure}

We call alternate network paths between a pair of {\it S-D} nodes \emph{distinct} -  as those with no overlapping edges between them. The number of distinct paths between a $S-D$ node pair depends on the degree of both the nodes, higher the degree of the nodes more the number of available paths between them. In RN and BAN, the number of distinct paths between an arbitrary $S-D$ pair cannot be deterministically estimated but in the regular network topologies we can obtain an exact estimation of the number of distinct paths and their respective lengths can also be determined in terms of the shortest graph path (SGP) length, $l_0$, between the two nodes. In TLN, SLN and HLN any two nodes, ignoring the nodes at the boundaries, are connected by 6, 4 and 3 distinct paths respectively. The length of the distinct paths between $S-D$ depends on the type the $D$ node around a particular $S$ node in a network topology. There are different types of $S-D$ node pair in the regular lattice networks, see Fig.~\ref{fig:network_topologies}. The lengths of the available distinct paths between $S-D$ node pair in the three regular lattice networks are listed in Fig.~\ref{fig:lengths_table}.



\section{Multipath entanglement purification}
\label{sec_three}
Here, we first describe the basics of Deutsch's protocol for entanglement pumping that we use for our MEP strategies, then we describe the analytical calculation that shows that the lengths of the network paths that are used for purification need to be comparable in an appropriate sense. Next, that a few paths are sufficient to approach the asymptotically obtainable concurrence and further that the order of purification is important. Finally, we devise two MEP strategies based on these observations.

\subsection{Deutsch's protocol for entanglement pumping }
We utilise Deutsch's protocol for entanglement purification also known as entanglement pumping \cite{purification_deutsch} to purify the states distributed along available paths between a pair of {\it S-D} nodes in a pairwise manner. In this multistep process, for a set of states $\rho_1,\rho_2,...,\rho_n$ used as input for the pumping process, each step utilises a pair of states to probabilistically obtain a state with higher concurrence. If a pumping step is successful the obtained state is then pumped using the next available state in the set till all states in the set are utilised. 

The Deutsch protocol is eminently suitable for MEP in quantum networks because it uses the entangled states along the multiple alternate network paths if they are available and as many are available. It is also efficient in terms of the number of memories required at any network node. We thus base our MEP strategies incorporating the Deutsch's protocol of entanglement purification.

\subsection{Comparable path lengths for useful purification}
\label{subsec:comparable_paths}
We now show that lengths of the network paths whose entangled states are purified need to be comparable for the pumping process to yield a concurrence of the output state which is larger than the more entangled of the two input states. To see this, consider that when two network states, $\rho_1,\rho_2$ defined as,
\begin{align}
\rho_1=(1-q)\ket{\phi^+}\bra{\phi^+}+q\frac{\mathbf{1_4}}{4},\nonumber\\
\rho_2=(1-q')\ket{\phi^+}\bra{\phi^+}+q'\frac{\mathbf{1_4}}{4},
\end{align}
are used in a single entanglement pumping step, the concurrence $C(\rho_{out})$ of the output state $\rho_{out}$ can be calculated to be \cite{purification_deutsch},
\begin{align}
C(\rho_{out})=\frac{q+q'-1+10(1-q)(1-q')}{1+2q+2q'+8(1-q)(1-q')}.
\end{align}
For the pumping to be useful we require, 
\begin{align}
C(\rho_{out})>\text{Max}(C(\rho_1),C(\rho_2)),
\end{align}
which in terms of the concurrence of the input states, $C(\rho_1)=C_1,C(\rho_2)=C_2$, and assuming, $C_1\geq C_2$, yield the following condition,
\begin{align}
 D(C_{1},C_{2}):=C_{1}-C_{2}< \frac{6}{h(C_{1})+6}+C_{1}-1,
\label{ineq_DeltaC}
\end{align}
where $h(C_{1}):= 2C_{1}(C_{1}+2)/(1+C_{1})(1-C_{1})$ is a monotonoically increasing convex function of $C_{1}$ with $h(0)=0, h(1)=\infty$. If we now take the absolute value of both sides of the above inequality we can upper bound the RHS as the sum of two terms, $|D(C_{1},C_{2})|<|6/\{h(C_{1})+6\}|+|C_{1}-1|$, both of which decrease as $C_{1}$ increases. This implies that the second pair of qubits in a different isotropic state needs to be closer in concurrence value as the first pair that we wish to purify gets more entangled. 

Now note that the average concurrences of the input states for the path purification are related to the path lengths, $l$, as,
\begin{align}
\overline{C_l}= 1-l\delta,~~~ l\ll (3/2)\ln 3/\delta,
\end{align}
a result which can be obtained by calculating the concurrence of a state obtained by entanglement swapping of the $l$ states along the path \cite{qn_topography}. The quantity, $\overline{C_l}$ is an average over different network paths of length $l$. This shows, that, on an average, the highest path concurrence is established between two nodes along the shortest graph path. Therefore, we ask if we can improve upon this value by purifying it using a pair established along an alternate path. Thus, we estimate the RHS of Ineq. (\ref{ineq_DeltaC}) for $C_{1}=\overline{C_l}$ and obtain $D(C_{1},C_{2})<l\delta$. The probability that the alternate path of length $(l+d)$ has concurrence within this value of that for the shortest-path (and  hence can be useful for purification) may then be estimated as,
\begin{align}
\text{Pr}(\text{path}~k_2~\text{can purify path} ~k_1)\leq 1+\frac{l-d}{d}\delta.
\label{ineq_pathpurify}
\end{align}
Note that the RHS is less than unity only when $d>l$. That is, as long as the alt-paths are not too long relative to the graph distance between {\it S-D} their entanglement resources can be usefully purified to provide a final state with a higher concurrence. In fact, the length of an alt-path that can be usefully combined increases with $l$ since paths with $d\in [1,2,...,l]$ can contribute with essentially no loss of probability. The gain in this process can be estimated to be,
\begin{align}
\Delta C_{1}=\frac{(l-d)\delta}{3}+O(\delta^2).
\label{gain_conc}
\end{align}
Thus Eq. (\ref{gain_conc}) together with Ineq. (\ref{ineq_pathpurify}) tells us that for $S-D$ separated by a graph distance $l$ all alt-paths with $0\leq d\leq (l-1)$ can provide a higher combined concurrence upon purification with high probability. Since there can be at most $k_{max}$ distinct alt-paths, the maximum gain in concurrence by combining alt-paths all assumed to be of the same length as the shortest-path, that is $d=0$, can be estimated to be,
\begin{align}
C_{k_1\oplus k_2 \oplus ...k_{max}}=1-(2/3)^{(k_{max}-1)}l\delta,
\label{conc_combined}
\end{align}
Note that in the region of $l\ll (1/\delta)$ that we consider, implying, short graph distances as compared to the maximum distance over which the concurrence can be shared, Eq. (\ref{conc_combined}) implies that $l\delta<<1$, therefore, even though the prefactor $(2/3)^{k_{max}-1}$ depends on $k_{max}$ which is the total number of purified paths, the effective post-purification concurrence has only a mild dependence on $k_{max}$. In other words combining only a few available network paths between a pair of {\it S-D} nodes is enough to approach the asymptotic value of the concurrence obtainable using entanglement pumping.

\subsection{Multipath purification strategies}
\label{subsec:mep_strategy}
An important guide for a multipath purification strategy can be obtained by analysing the entanglement purification of states distributed along $k_{max}$ distinct paths between $S$ and $D$ with lengths $l_0, l_1,..., l_{k_{max}-1}$ in a homogeneous network where each edge is distributed with entangled states having concurrence $c$. This requires us to define and adopt specific pumping strategies which we now proceed to do. A specific pumping strategy, $\mathcal{S}_P(0,1,2,...,(k_{max}-1))$, for the pumping steps of the $(k_{max}-1)$ states along the different paths can be as follows:  $\mathcal{S}_P(0,1,2,...,(k_{max}-1)):=l_0\oplus l_1\oplus l_2...\oplus l_{(k_{max}-1)}$ which is taken to mean that the entanglement distributed along the two paths of lengths $l_0$ and $l_1$ is purified first. Then, the (probabilistically) obtained purified entangled state in the previous step is pumped with the state distributed along the path of length $l_2$ and so on till the last path of length $l_{(k_{max}-1)}$ is used in the $(k_{max}-1)$'th pumping step. Therefore, schematically,
\begin{align}
\mathcal{S}_P(0,1,2,3,...,k_{max}-1):=\underbrace{\underbrace{\underbrace{\underbrace{l_0\oplus l_1}_{\text{Pump 1}}\oplus l_2}_\text{{Pump 2}}\oplus l_3}_\text{Pump 3}...\oplus l_{(k_{max}-1)}}_\text{Pump $(k_{max}-1)$},
\end{align}
is a strategy which sequentially pumps previously purified states with those obtained along paths of length $l_i,0\leq i\leq (k_{max}-1)$. 

The concurrence of the state obtained at the end of the strategy $\mathcal{S}_P(0,1,2)$ is given by,
\begin{align}
    &C^{(k=3)}(l_0,l_1,l_2)=\nonumber\\
    &\frac{1-\frac{1}{6}f_1(1-c)+\frac{1}{18}f_2(1-c)^2-\frac{7}{54}f_4(1-c)^3}{1-\frac{1}{6}(f_1-2l_2)(1-c)+\frac{1}{18}f_3(1-c)^2-\frac{4}{27}f_4(1-c)^3}
    \label{combined_conc}
\end{align}
Here, $f_1(l_0,l_1,l_2)=(3l_0+3l_1+4l_2)$, $f_2(l_0,l_1,l_2)=(4l_0l_1+5l_1l_2+5l_0l_2)$, $f_3(l_0,l_1,l_2)=(6l_0l_1+4l_1l_2+4l_0l_2)$ and $f_4(l_0,l_1,l_2)=l_0l_1l_2$. Now, notice that the functions $f_1(l_0,l_1,l_2)$, $f_2(l_0,l_1,l_2)$, $f_3(l_0,l_1,l_2)$ are not invariant under the exchange of the path length indices $i$ for all $i=0,1,2$. Therefore, we conclude that the resultant concurrence after the entanglement pumping steps depends on the order in which the different paths are combined. 

An important corollary, critical for guiding an efficient multipath purification strategy, is obtained by noticing that for $(1-c)\to0$ in Eq. (\ref{combined_conc}), we obtain 
\begin{align}
C^{(k=3)}\approx 1-\frac{l_2}{3}(1-c),
\end{align}
which implies that the resultant concurrence obtained via the strategy $\mathcal{S}_P(0,1,2)$ is mainly determined by the length of the path that we use in the final entanglement pumping step. For a higher resultant concurrence one must pump using the entanglement along the shortest-path $l_0$ in the scenario when the mean concurrence in the network is close to the maximum. 

We propose two different purification strategies based on the insights from our previous discussion : Shortest-path first multipath entanglement purification (SPF-MEP) and Shortest-path last multipath entanglement purification (SPL-MEP). We now desribe the two strategies and show analytical description of the pumped path concurrences according to the two respective strategies for homogeneous networks based on regular lattices.

\subsubsection{Shortest-path first-MEP (SPF-MEP)}
This strategy considers the purification of the available distinct paths according to the pumping strategy $\mathcal{S}_P(0,1,2)$ where the SGP is pumped at the initial pumping step. 

We present here an analytical description of the average pumped path concurrence between $S-D$ pairs in homogeneous networks based on the three regular lattice structures according to SPF-MEP. These are obtained by averaging over different types of $D$ nodes at a particular $l_0$. The average pumped path concurrence in homogeneous TLN, SLN and HLN according to SPF-MEP are given by three different functions,$\overline{C^{SPF}_{TLN}}$, $\overline{C^{SPF}_{SLN}}$ and $\overline{C^{SPF}_{HLN}}$ respectively, for the three different network topologies. The functions for the three different network structures are given by,
\begin{align}
    \overline{C^{SPF}_{TLN}} =
    \begin{cases}
        C^{(k=3)}(1,2,2), & \text{for } l_0 = 1\\
        \\
        \begin{aligned}
            &\frac{1}{l_0}C^{(k=3)}(l_0,l_0+1,l_0+1) \\
            &+ \frac{l_0-1}{l_0}C^{(k=3)}(l_0,l_0,l_0+2),
        \end{aligned} & \text{for } l_0 \neq 1
    \end{cases}
    \label{spf_eqns}
\end{align}
\begin{align}
    \overline{C^{SPF}_{SLN}} =
    \begin{cases}
        \begin{aligned}
            &\frac{1}{l_0}C^{(k=3)}(l_0,l_0,l_0+4) \\
            &+ \frac{l_0-1}{l_0}C^{(k=3)}(l_0,l_0+2,l_0+2),
        \end{aligned} & \text{for } l_0 =\text{even} \\
        \\ 
        C^{(k=3)}(l_0,l_0+2,l_0+2),& \text{for } l_0 =\text{odd}
    \end{cases}
\end{align}
\begin{align}
    \overline{C^{SPF}_{HLN}} =
    \begin{cases}
        C^{(k=3)}(l_0,l_0+2,l_0+6), & \text{for } l_0 = \text{even}\\
        \\
        \begin{aligned}
            &\frac{l_0+1}{2l_0}C^{(k=3)}(l_0,l_0+2,l_0+6) \\
            &+ \frac{l_0-1}{2l_0}C^{(k=3)}(l_0,l_0,l_0+8),
        \end{aligned} & \text{for } l_0=\text{odd}
    \end{cases}
\end{align}

\subsubsection{Shortest-path last - MEP (SPL-MEP)}
In this strategy, we pump the longer paths at the initial step of purification and then the SGP is pumped at the final pumping step according to the pumping strategy $\mathcal{S}_P(2,1,0)$.

The analytical form of the average pumped path concurrence according to SPL-MEP for homogeneous networks based on three regular lattice topologies are given  by the functions $\overline{C^{SPL}_{TLN}}$, $\overline{C^{SPL}_{SLN}}$ and $\overline{C^{SPL}_{HLN}}$ for TLN, SLN and HLN respectively. The functions are given below.
\begin{align}
    \overline{C^{SPL}_{TLN}} =
    \begin{cases}
        C^{(k=3)}(2,2,1), & \text{for } l_0 = 1\\
        \\
        \begin{aligned}
            &\frac{1}{l_0}C^{(k=3)}(l_0+1,l_0+1,l_0) \\
            &+ \frac{l_0-1}{l_0}C^{(k=3)}(l_0+2,l_0,l_0),
        \end{aligned} & \text{for } l_0 \neq 1
    \end{cases}
\end{align}
\begin{align}
    \overline{C^{SPL}_{SLN}} =
    \begin{cases}
        \begin{aligned}
            &\frac{1}{l_0}C^{(k=3)}(l_0+4,l_0,l_0) \\
            &+ \frac{l_0-1}{l_0}C^{(k=3)}(l_0+2,l_0+2,l_0),
        \end{aligned} & \text{for } l_0 =\text{even} \\
        \\ 
        C^{(k=3)}(l_0+2,l_0+2,l_0),& \text{for } l_0 =\text{odd}
    \end{cases}
\end{align}
\begin{align}
    \overline{C^{SPL}_{HLN}} =
    \begin{cases}
        C^{(k=3)}(l_0+6,l_0+2,l_0), & \text{for } l_0 = \text{even}\\
        \\
        \begin{aligned}
            &\frac{l_0+1}{2l_0}C^{(k=3)}(l_0+6,l_0+2,l_0) \\
            &+ \frac{l_0-1}{2l_0}C^{(k=3)}(l_0+8,l_0,l_0),
        \end{aligned} & \text{for } l_0=\text{odd}
    \end{cases}
    \label{spl_eqns}
\end{align}

Note, Eqs. (\ref{spf_eqns}-\ref{spl_eqns}) provide the quantities that we can estimate from numerical simulations on different network topologies that can indicate the effectivness of MEP on such networks. This is the content of our next section.

\section{Numerical Results}
\label{sec_four}
We now describe the results of numerical simulations performed on networks that were implemented in Python using the NetworkX package. The quantum networks considered had $10^4$ nodes and about $5\times10^4$ edges. MEP strategies were tested on network realisations with mean edge-concurrence of $\overline{c_e}=0.98$ and a minimum of $c_{min}=0.97$ and a maximum of $c_{max}=0.99$. Pairs of {\it S-D} nodes were sampled with varying shortest graph path distances, $l_0$, between them. Due to graph theoretic properties of random networks enough {\it S-D} samples were obtained only for $1\leq l_0\leq 6$ since their small-world property implies that arbitrary pairs of nodes are very likely to be at most separated by $l_0=6$. At every value of $1\leq l_0\leq 6$, $10^2$ source nodes $S$ in relation to $10^2$ destination nodes $D$ were sampled. Comparison of the average concurrence scaling with graph distance between the source and destination nodes reveal important features of the MEP strategies in relation to network topologies.

\subsection{General observations for multipath entanglement purification in quantum networks}
We now highlight some general empirical inferences about MEP in quantum networks that can be drawn from numerical results obtained on network samples obtained using the procedure above. 

\begin{figure}
    \centering
    \includegraphics[width=\columnwidth]{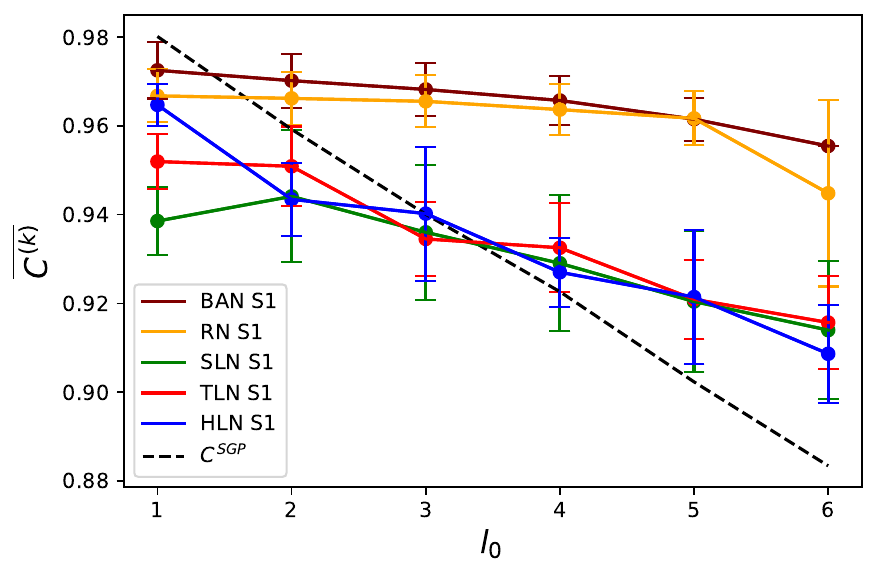}
    \caption{Scaling of the effective concurrence $\overline{C^{(k)}}$ using multipath entanglement purification between {\it S-D} nodes vs their shortest-path separation $l_0$ on inhomogenous networks with various topologies using the SPF-MEP strategy. Each network has a size of about $10^4$ nodes and $5\times10^4$ edges. For Random network (RN) and Barabasi-Albert network (BAN) 3 paths are combined but for other three regular networks all paths available are combined, i.e., 3 for Hexagonal (HLN), 4 for square (SLN) and 6 for triangular (TLN). The black dotted line shows the average concurrence obtained along the shortest graph paths between two nodes.}
    \label{fig:fig_1}
\end{figure}

The results of implementing the two strategies SPF-MEP and SPL-MEP are shown in the two figures, Fig. \ref{fig:fig_1} and \ref{fig:fig_2}. . While in Fig. \ref{fig:fig_1} only results from the SPF-MEP are shown, in Fig. \ref{fig:fig_2} results from both the SPF and SPL strategies are shown to allow a visual and logical comparison.

Focusing on Fig. \ref{fig:fig_1}, we see the scaling of the effective concurrence, $\overline{C^{(k)}}$, using MEP of $k$ number of paths between {\it S-D} nodes vs their shortest-path separation $l_0$ for inhomogeneous networks with various topologies. For these results, $k=3,4,6$ number of paths were used for the regular network topologies which correspond to the node degree for HLN, SLN and TLN respectively. For the random (RN) and scale-free (BAN) topologies the maximum number of combined paths were capped at $k=3$. Several interesting observations can be made from this plot as follows.

First, notice that at the shortest-path length, $l_0=1$, there is no correlation between the network's average node degree and the effective concurrence. The hexagonal lattice (in Blue) has nodes with degree $3$, yet combining only the three available paths in this case yields a better post-purified concurrence than for the triangular and square lattices with degree 6 and 4 respectively. The random (Yellow) and scale-free network (Brown) which do not have a fixed node degree seem to perform better at this value of $l_0$ - a trend which persists for larger graph distances. This implies that,
\begin{itemize}
\item {\it The effectiveness of MEP strategies are not (strongly) determined by the number of paths purified.}
\end{itemize}

Next, from Fig. \ref{fig:fig_1}  notice that for the same value of $l_0=1$, MEP is infact detrimental in the SPF case because the concurrence value obtained along the shortest graph path without purification (black dotted line) is higher than all the cases where MEP was implemented. We explain this behavior by refering to the discussion earlier where the alternate paths needed to have comparable lengths for MEP to be useful. There, we had shown that for an alternative path of length, $(l_0+d)$, between a pair of {\it S-D} nodes separated by a graph distance of, $l_0$, the value of, $d$, needed to be within, $0\leq d \leq (l_0-1)=0$. This motivates our next observation that,
\begin{itemize}
\item {\it MEP based on arbitrary strategies is not necessarily useful for very short graph distances when the average edge-concurrence in the network is high. The crossover distance depends on the network topology.}
\end{itemize}

Finally, from Fig. \ref{fig:fig_1} and \ref{fig:fig_2} notice that MEP gives an advantage relative to entanglement connections only along the single shortest graph path for $l_0$ larger than the crossover graph distances between {\it S-D} nodes. This crossover graph distance is the value of $l_0$ at the point where  the solid plot lines and the Black dotted line intersect. For the random and scale-free network this crossover occurs immediately beyond the shortest graph distances, whereas for the regular networks this cross over is at $l_0=3$. We collect this observation as the inference that,
\begin{itemize}
\item {\it MEP is potentially useful for graph distances beyond a crossover shortest-path distance between source-destination nodes.}
\end{itemize}

\subsection{ Comparison of Shortest-path first and Shortest-path last MEP strategies on a quantum network}
\begin{figure}
    \centering
    \includegraphics[width=1.1\columnwidth]{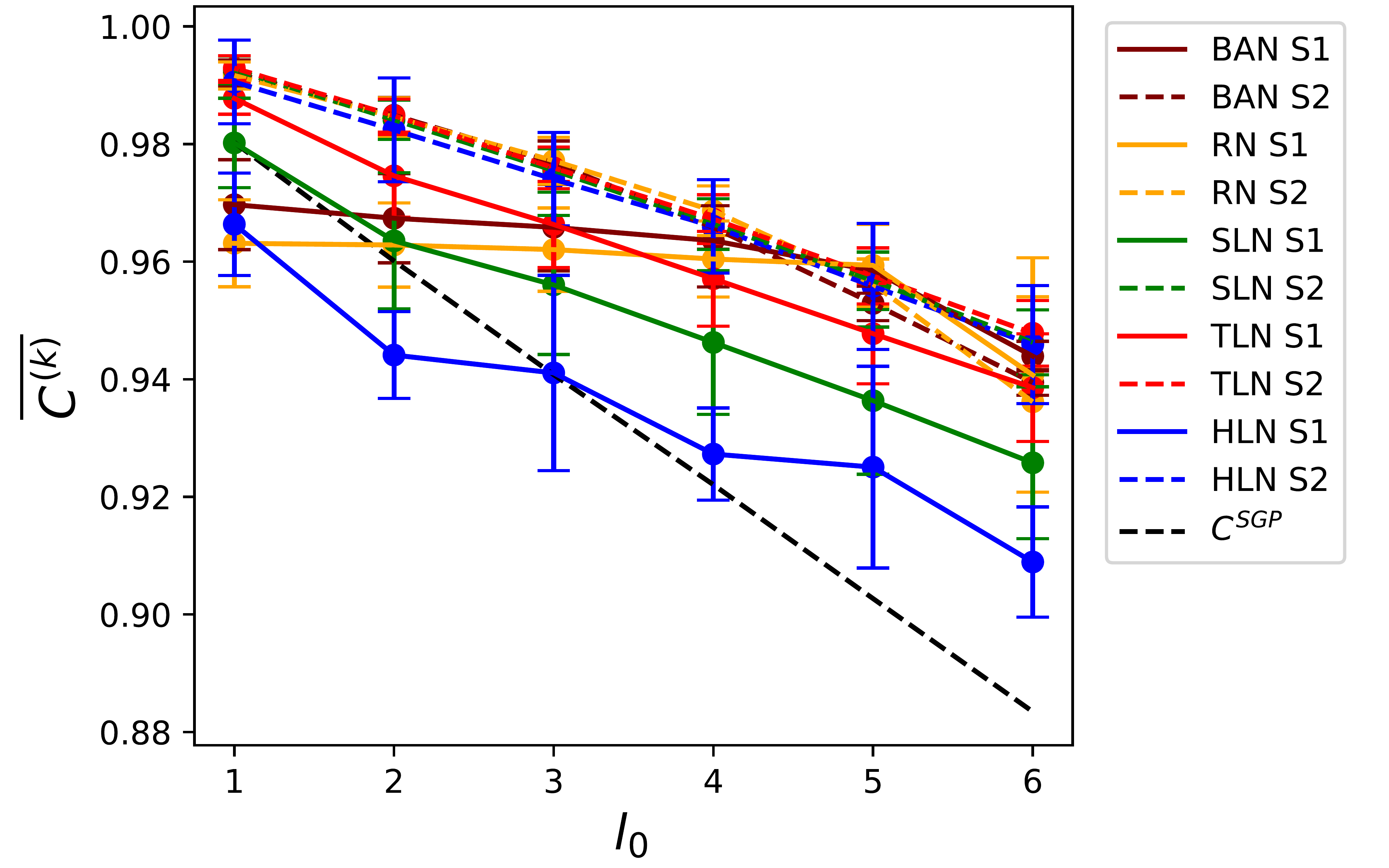}
    \caption{Scaling of the effective concurrence $\overline{C^{(k)}}$ using multipath entanglement purification between {\it S-D} nodes vs their shortest-path separation $l_0$ on inhomogenous networks with various topologies using both the SPF-MEP (S1) and SPL-MEP (S2) strategies. Each network has a size of about $10^4$ nodes and $5\times10^4$ edges. For all the networks only 3 paths between any {\it S-D} node pair are combined.}
    \label{fig:fig_2}
\end{figure}

The results of implementing both the SPF-MEP and SPL-MEP are shown in Fig. (\ref{fig:fig_2}). The figure shows the scaling of the effective concurrence, $\overline{C^{(k)}}$, using MEP of $k=3$ (fixed) number of paths between {\it S-D} nodes vs their shortest-path separation $l_0$ for inhomogeneous networks with various topologies. For the random (RN) and scale-free (BAN) topologies all available paths were combined with the maximum number of paths capped at $k=3$. Results of SPF-MEP are shown using solid color lines while those for SPL-MEP are shown using dotted lines. Several interesting observations about the effectiveness of the two strategies in comparison can be made from this figure as follows.

First, notice from Fig. \ref{fig:fig_2} that the scaling of the effective multipath purified concurrence, $\overline{C^{(k)}}(l_0)$, with the distance, $l_0$, between the {\it S-D} nodes is significantly determined by the network topology. For the SPF-MEP, while the regular network topologies, TLN, SLN, HLN shown in solid Red, Green and Blue respectively have practically the same scaling with, $l_0$, the random and scale-free graphs shown in Solid Yellow and Brown have a very different scaling with a plateau for a range of distances, $1\leq l_0\leq 4$. Second, in the same figure notice that for the SPL-MEP strategy, shown in dotted lines, the effective concurrences for all network topologies at all graph distances roughly coincide and show the same scaling with the distance, $l_0$, between the {\it S-D} nodes. In other words, this strategy eliminates the dependence of the effectiveness of MEP on the network topology.  This implies that,
\begin{itemize}
\item {\it Network topology is a crucial factor in determining the effectiveness of a given MEP strategy.}
\end{itemize}

Next, notice from Fig. \ref{fig:fig_2} that the effective post-purified concurrence for the SPL-MEP strategy (dotted lines) for all network topologies is better than the no-purification strategy and the SPF-MEP strategy for all graph distances $l_0$. In particular, for $l_0=1$ while the SPF-MEP performs badly the SPL-MEP does very well improving the strength of the entanglement connection also for short distances even with average high edge-concurrence values. This implies that,
\begin{itemize}
\item {\it Effective multipath entanglement purification depends significantly on the choice of the strategy. A careful strategy choice can make MEP effective for all network regions.}
\end{itemize}

Finally, notice from Fig. \ref{fig:fig_2} that the crossover point for MEP to perform better than no-purification strategies, is here at $l_0=0$. This means that even for the shortest graph distances MEP using the shortest-path last strategy, SPL-MEP, is more effective at establishing entanglement connections than only the no-purification single shortest-path entanglement connections. This leads to our final observation that,
\begin{itemize}
\item {\it MEP strategies can be chosen to enhance the entanglement connections at all distance scales in a quantum network.}
\end{itemize}

\section{Discussion and conclusions}
\label{sec_five}
In this work we have shown that multipath entanglement purification can enhance the entanglement connection between source-destination node pairs in a quantum network leading to potential implementation of quantum information processing tasks over larger network regions. By analysing the Deutsch's purification protocol for entangled quantum states distributed via multiple alternate paths between a pair of source-destination nodes we obtained design criteria that can guide efficient MEP strategies and two such strategies were evaluated based on their performance for various network topologies.

Our results indicate that including the shortest-path last in the pumping step statistically provides better resultant multipath purified concurrence between node pairs. Importantly, only a few paths need to be purified when the average network entanglement is large. The random and scale-free network topologies provide the most favorable scaling for  the concurrence as a function of the distance between the {\it S-D} pair. From a practical point of view, our results indicate that multipath entanglement purification can be an important network protocol to increase the viability of quantum networks for distributed quantum information processing tasks. Although entanglement purification \cite{exptl_purif_pan,exptl_purif_huang} is a more complex quantum operation relative to entanglement swapping \cite{entswap_conc,kirby-swapping,exptl_swapping} and will require significantly better quantum networking hardware resources, in networks \cite{azuma2023quantum} of quantum repeaters \cite{Santra_20191,Santra_20192} purification is one of the basic network operations. In such networks, therefore, multipath entanglement purification can be a natural addition to the set of network operations that can significantly enhance the set of nodes over which bipartite entanglement based quantum information tasks can be performed. Multipath entanglement purification strategies would also need to be integrated with routing protocols in quantum networks \cite{ent-routing,leone2021qunet} in order to harness the complex network structure of large scale networks towards long-distance entanglement distribution. Therefore, various multipath entanglement purification strategies need to be studied in conjunction with multipath entanglement routing \cite{ent-routing,leone2021qunet} and multi-commodity flow \cite{azuma2023networking} protocols with the goal of efficient quantum network design.

\section*{Acknowledgment}
We acknowledge funding from DST, Govt. of India through the SERB grant MTR/2022/000389, IITB TRUST Labs grant DO/2023-SBST002-007 and the IITB seed funding.

\bibliographystyle{IEEEtran}

\end{document}